\documentclass[a4paper,11pt]{article}
\pdfoutput=1 % if your are submitting a pdflatex (i.e. if you have
\usepackage{jheppub,bm} % for details on the use of the package, please
\usepackage[T1]{fontenc} % if needed
\newcommand{\nn}{\nonumber}

\title{Post-Newtonian dynamics basing on Mathisson-Papapetrou equations with Corinaldesi-Papapetrou condition in Kerr spacetime}

\author[a,b]{Chunhua Jiang}
\author[a,1]{and Wenbin Lin\note{Corresponding author.}}

\affiliation[a]{School of Physical Science and Technology, Southwest Jiaotong University,\\111 North 1 Section, Erhuan Road, Chengdu, China}
\affiliation[b]{School of Mathematics and Physics, University of South China, \\28 Changsheng West Road, Hengyang, China}

\emailAdd{jiangchunhua@my.swjtu.edu.cn}
\emailAdd{wl@swjtu.edu.cn}

\abstract{We derive the post-Newtonian dynamics for a spinning body with Corinaldesi-Papapetrou spin supplementary condition in Kerr spacetime. Both the equations of motion for the center-of-mass of body and the spin evolution are obtained. For the non-relativistic case, our calculations show that the magnitude of spin measured in the rest frame of the body's center-of-mass does not change with time, though the center-of-mass does not move along the geodesic. Moreover, we find that the effects of the spin-orbit and spin-spin couplings will be suppressed by the Lorentz factor when the body has a relativistic velocity.}

\begin{document}
\maketitle
\flushbottom

\section{Introduction} \label{sec:intro}

The dynamics of the point particle in Kerr field has been presented in many classical textbooks \cite{Weinberg1972,MTW1973,Chandrasekhar1983,Landau1971}. More further studies can be found in recent papers \cite{Dymnikova1986,Pugliese2011,JiangLin2014}. The dynamics of the extended body is much more complex, because the internal structure of the body usually subjects to the gravitational field and makes the body follow a nongeodesic trajectory.

The first attempt to explore the dynamics of a spinning body in the curved spacetime is performed by Mathisson \cite{Mathisson1937}, who shows there is an interaction term coupling the body's spin tensor to the Riemann curvature tensor in the equations of motion.
%{\color{red}Papapetrou \cite{Papapetrou1951} derives the general equations of motion for the spinning body with respect to the body's proper time, in the pole-dipole approximation. These equations are commonly referred to as Mathisson-Papapetrou (MP) equations, and the spinning body under the pole-dipole approximation is also called as the spinning particle.}
Papapetrou \cite{Papapetrou1951} derives the equations of motion for the extended body with respect to the proper time, and demonstrates that the body moves along the geodesic in the single-pole approximation and the spin effect appearing in the pole-dipole approximation makes the worldline of the body deviate from the geodesic. The extended body under the pole-dipole approximation is also called as the spinning particle, and the general equations of motion for the spinning particle are commonly referred to as Mathisson-Papapetrou (MP) equations.

MP equations are not a closed system, and the spin supplementary condition (SSC) needs to be imposed to fix the representative worldline of the extended body. The commonly-used SSCs are Corinaldesi-Papapetrou (CP)~\cite{PapapetrouCorinaldesi1951}, Pirani (P)~\cite{Pirani1956} and Tulczyjew (T)~\cite{Tulczyjew1959} conditions. CP condition chooses the center-of-mass of the body in the rest frame of the gravitational source as the representative point. P and T conditions choose the center-of-mass of the body in its own rest frame and in the zero 3-momentum frame as the representative point, respectively.
Although different SSCs lead to different solutions of MP equations for the same physical motion, the choice of SSCs still remains a not well understood problem~\cite{Costa2012,Costa2014}.

The motion of the spinning particle in the field of black hole has been studied by many researchers~\cite{Barker1974,Barker1979,Schiff1960a,Kyrian2007,Bini2004,Bini2006,Rasband1973,Han2008,Bini2000,Chicone2005,Mashhoon2006,Plyatsko2005,Suzuki1997,Plyatsko2012,Plyatsko2013,Plyatsko2008,Hartl2003,Apostolatos1996,Semerak1999}. Specifically,
Schiff derives the equations for the spin evolution in Schwarzschild field under CP and P conditions respectively~\cite{Schiff1960a}.
Barker and O'Connell study the nongeodesic motion of the spinning particle in Schwarzschild field under CP and P conditions, and demonstrate that  the equations of motion under these two conditions are consistent with each other by shifting in the center of mass~\cite{Barker1974}.
Bini, Gemelli, and Ruffini invesigate the behaviour of charged spinning particle moving along circular orbits in the equatorial plane of Reissner-Nordstr\"{o}m spacetime under CP, P and T conditions respectively~\cite{Bini2000}. Kyrian and Semer\'{a}k solve MP equations with various SSCs in Kerr field numerically, and show that the trajectories of the representative point of the spinning particle under different SSCs are different \cite{Kyrian2007}.
Plyatskon and Fenyk investigate the highly relativistic circular orbits of a spinning particle in Schwarzschild field under
P and T conditions~\cite{Plyatsko2012}, and the different cases of the spin orientation and the direction of the particle's orbital motion on the highly relativistic circular orbits in the equatorial plane of Kerr source under P condition~\cite{Plyatsko2013}.
Recently, Gralla and Wald also derive the gravitational self-force of the spinning particle in Schwarzschild-de Sitter spacetime under CP condition~\cite{GrallaWald2008}.
In this work we study the dynamics of the spinning particle in Kerr field under CP condition, which so far has not been explored. Basing on the harmonic metric for Kerr spacetime~\cite{JiangLin2014}, we derive the post-Newtonian equations of motion for the spinning particle including the spin evolution, to the next-to-leading terms of the spin-curvature coupling.

The content of this paper is arranged as follows. Section \ref{sec2} briefly introduces MP equations, and gives the equation of motion for the spin under the CP condition in the post-Newtonian approximation. In Section \ref{sec3} we derive the equations for the particle's motion and spin evolution in Kerr spacetime, and calculate the spin precession measured in the rest frame of the particle for the non-relativistic case. Summary is given in Section \ref{conclusion}.

\section{\label{sec2}Mathisson-Papapetrou equations}

A spinning particle deviates from the geodesic motion and moves along an orbit described by MP equations of motion \cite{Papapetrou1951}
\begin{eqnarray}
&&\frac{D}{Ds}\Big(m u^\alpha + u_\beta\frac{DS^{\alpha\beta}}{Ds}\Big)+\frac{1}{2}S^{\rho\lambda} u^\sigma R_{\lambda\sigma\rho}^\alpha =0~,\label{motionEq:a}\\
&&\frac{DS^{\alpha\beta}}{Ds} + u^\alpha u_\rho \frac{DS^{\beta\rho}}{Ds} - u^\beta u_\rho \frac{DS^{\alpha\rho}}{Ds} = 0~, \label{spinEq:a}
\end{eqnarray}
where $u^\alpha\equiv dx^\alpha/ds$ and $S^{\alpha\beta}$ are the particle's contravariant 4-velocity vector and spin tensor, respectively. $m$ denotes the kinematical rest mass of the spinning particle, and it reduces to the usual rest mass when the spin vanishes. $D/Ds$ denotes  the covariant derivative with respect to the particle's proper time $s$. $R_{\lambda\sigma\rho}^\alpha$ is the Riemann curvature tensor corresponding to the background on which the particle moves. For the spinless particle one can easily verify that the above trajectory Eq.~(\ref{motionEq:a}) reduces to the usual geodesic equation. Here and in the following, we use the geometrized units ($c=G=1$) and the metric signature ($---+$), with Greek indices running from 1--4 and Latin indices from 1--3.

The kinematical rest mass $m$ of the spinning particle may not be a constant, and its evolution is described by~\cite{Semerak1999}
\begin{equation}\label{mparameter}
    \frac{Dm}{Ds} = \frac{D(p^\alpha u_\alpha)}{Ds} = \frac{Du_\alpha}{Ds} u_\beta \frac{DS^{\alpha\beta}}{Ds}.
\end{equation}
where $p^\alpha$ is the contravariant 4-momentum vector of the spinning particle
    \[p^\alpha \equiv m u^\alpha + u_\beta\frac{DS^{\alpha\beta}}{Ds}~.\]

Substituting Eq.~(\ref{mparameter}) into Eq.~(\ref{motionEq:a}), we have
\begin{equation}\label{motionEq:b}
\hskip -1cm    m \frac{Du^\alpha}{Ds} = -\frac{Du_\mu}{Ds}\frac{DS^{\mu\nu}}{Ds}u_\nu u^\alpha - \frac{D}{Ds}\Big(u_\beta \frac{DS^{\alpha\beta}}{Ds}\Big) - S^{\mu\nu}u^\sigma(\Gamma_{\nu\sigma,\mu}^\alpha + \Gamma_{\mu\rho}^\alpha \Gamma_{\nu\sigma}^\rho)~.
\end{equation}
General speaking, the acceleration of the spinning particle is given by
\begin{equation}\label{eq:acceleration}
  \frac{dv^i}{dt} = \frac{1}{(u^0)^2}\left(\frac{du^i}{ds} - v^i \frac{du^4}{ds}\right) = a_{\rm geo}^i + a_{\rm spin}^i~,
\end{equation}
where $v^i = dx^i/dt=u^i/u^4$ is the velocity of the particle, and
\begin{eqnarray}
  &&a_{\rm geo}^i = -\frac{u^\mu u^\nu}{(u^4)^2}(\Gamma_{\mu\nu}^i - v^i\Gamma_{\mu\nu}^4)~, \label{eq:geo-acceleration}\\
  &&a_{\rm spin}^i = \frac{1}{(u^4)^2}\left(\frac{Du^i}{Ds} - v^i \frac{Du^4}{Ds}\right)~, \label{eq:spin-acceleration}
\end{eqnarray}
in which $a_{\rm geo}^i$ and $a_{\rm spin}^i$, respectively, denote the acceleration caused by the curvature of spacetime and the coupling of spin-curvature.

For the evolution of spin, since $S^{\alpha\beta}$ is an anti-symmetric tensor, it has six independent components. However, Eq.~(\ref{spinEq:a}) only has three independent equations, which can be re-written as \cite{Papapetrou1951}
\begin{equation}\label{spinEq:b}
    \frac{DS^{\alpha\beta}}{Ds} = \frac{u^\beta}{u^4} \frac{DS^{\alpha4}}{Ds} -\frac{u^\alpha}{u^4} \frac{DS^{\beta4}}{Ds}~,
\end{equation}
therefore, some supplementary conditions have to be imposed. In this work we only consider the Corinaldesi-Papapetrou condition \cite{PapapetrouCorinaldesi1951}
\begin{equation}\label{PCcondition}
    S^{i4} = 0~,
\end{equation}
which implies that $x^{\alpha}$ is the center-of-mass of the spinning particle, and thus makes Eq.~(\ref{motionEq:a}) describe the trajectory for the particle's center-of-mass in the curved spacetime \cite{Schiff1960a}.

In the post-Newtonian approximation, the affine connection $\Gamma_{\mu\nu}^\lambda$ can be expanded as follows \cite{Weinberg1972}
\begin{eqnarray}
&&\Gamma_{\mu\nu}^\lambda = \stackrel{2~~~}{\Gamma_{\mu\nu}^\lambda} + \stackrel{4~~~}{\Gamma_{\mu\nu}^\lambda} + \cdots \quad (\mbox{for}~~ \Gamma_{44}^i, \Gamma_{jk}^i, \Gamma_{4i}^4), \nn\\
&&\Gamma_{\mu\nu}^\lambda = \stackrel{3~~~}{\Gamma_{\mu\nu}^\lambda} + \stackrel{5~~~}{\Gamma_{\mu\nu}^\lambda} + \cdots \quad (\mbox{for}~~ \Gamma_{4j}^i, \Gamma_{44}^4, \Gamma_{ij}^4), \label{affineconnect}
\end{eqnarray}
where the symbol $\stackrel{N~~~}{\Gamma_{\mu\nu}^\lambda}$ denotes the term in $\Gamma_{\mu\nu}^\lambda$ of order $\bar{v}^N/\bar{r}$, with $\bar{v}$ and $\bar{r}$ representing respectively the typical values of velocity and distance in a non-relativistic system.

Substituting Eqs.~(\ref{PCcondition}) and (\ref{affineconnect}) into Eq.~(\ref{spinEq:b}), keeping the coefficient of $S^{ij}$ on the right-hand side of Eq.~(\ref{spinEq:b}) to order $\bar{v}^3/\bar{r}$, we have
\begin{eqnarray}
    &&\frac{DS^{i4}}{Ds} = \Big(\stackrel{3~~}{\Gamma_{kj}^4} u^j + \stackrel{2~~}{\Gamma_{k4}^4} u^4\Big)S^{ik}~,  \label{s4i} \\
    &&\frac{DS^{ij}}{Ds} = \Big(\stackrel{3~~}{\Gamma_{kl}^4} \frac{u^l}{u^4} +\stackrel{2~~}{\Gamma_{k4}^4}\Big) (u^j S^{ik}-u^i S^{jk})~. \label{sij}
\end{eqnarray}

Eqs.~(\ref{motionEq:b}), (\ref{s4i}), and (\ref{sij}) are the basis for us to obtain the equation of motion for the center-of-mass of the spinning particle, as well as the spin evolution.

\section{Equations of motion in Kerr spacetime}\label{sec3}

The spacetime for a constantly rotating gravitational source is referred to as Kerr spacetime. In the harmonic coordinates, Kerr metric can be written as \cite{JiangLin2014,LinJiang2014}
\begin{eqnarray}
ds^{2}&=&dt^{2}-\frac{r^2(r+M)^2+a^2z^2}{\big(r^2+\frac{a^2}{r^2}z^2\big)^2}\Bigg[\frac{\big(\bm{x}\cdot d\bm{x}+\frac{a^2}{r^2}z dz\big)^{2}}{r^2+a^2-M^2}+\frac{z^2}{r^2} \frac{\big(\bm{x}\cdot d\bm{x}-\frac{r^2}{z} d z\big)^2}{r^2-z^2}\Bigg]\nn\\
&& -\frac{(r+M)^{2}+a^{2}}{r^2-z^2}\Bigg[\frac{a r^2M^2(r^2-z^2)\big(\bm{x}\cdot d\bm{x}+\frac{a^2}{r^2}z dz\big)}{(r^2+a^2-M^2)(r^2+a^2)(r^4+a^2 z^2)}+\frac{r(y dx-x dy)}{r^2+a^2}\Bigg]^{2} \nn\\
&& -\frac{2M(r+M)}{(r+M)^2+\frac{a^2}{r^2}z^2}\!\Bigg[\!\frac{a^2rM^2(r^2-z^2)\big(\bm{x}\cdot d\bm{x}+\frac{a^2}{r^2}z dz\big)}{(r^2+a^2\!-\!M^2)\!(r^2+a^2)\!(r^4+a^2z^2)}\!+\!\frac{a(y dx-x dy)}{r^2+a^2}\!+\!dt\!\Bigg]^{2}\!.~~ \label{HarmonicKN}
\end{eqnarray}
where $M$ and $a$ denote the mass and the angular momentum per mass for the gravitational source, respectively. The angular momentum is assumed to be codirectional with the positive direction of $z$-axis. $\bm{x} \equiv$ ($x$, $y$, $z)$, $\bm{x}\cdot d\bm{x}\equiv x dy + y dy + z dz$, and $(x^2+y^2)/(r^2+a^2)+z^2/r^2=1$. Expanding the coefficients of the metric Eq.~\eqref{HarmonicKN} in powers of $1/r$ to next-to-next-to leading order, we have
\begin{eqnarray*}
g_{ij}&=&-\Big(1-2\phi + \phi^2\Big)\delta_{ij} - \frac{\phi^2x^i x^j}{r^2}~,\\
g_{i4}&=&-(1+\phi)\zeta_i~, \\
g_{44}&=&1+2\phi+2\phi^2~,
\end{eqnarray*}
where $\phi \equiv -M/r$ is Newtonian potential, $\bm{\zeta}\equiv 2 a M (\bm{x}\times\bm{e}_{3})/r^3$ denotes the vector potential due to the gravitational source's rotation, with $\bm{e}_{3}$ being the unit vector of $z$-axis.

The corresponding affine connection which are needed in the following calculations can be written as follows
\begin{eqnarray}
&&\stackrel{2~~~}{\Gamma_{jk}^i} = -\frac{M}{r^3}(x^k \delta_{ij}+x^j \delta_{ik}-x^i \delta_{jk}), \nn\\
&&\stackrel{2~~~}{\Gamma_{44}^i} = \stackrel{2~~~}{\Gamma_{i4}^4} = \frac{M}{r^3}x^i, ~
\stackrel{3~~~}{\Gamma_{j4}^i} = \frac{1}{2}(\zeta_{i,j}-\zeta_{j,i}),\nn\\
&&\stackrel{3~~~}{\Gamma_{jk}^4} = -\frac12 (\zeta_{j,k} + \zeta_{k,j}), ~
\stackrel{4~~~}{\Gamma_{j4}^4} = 0, \label{Christoffel}\\
&&\stackrel{4~~~}{\Gamma_{jk}^i} = \frac{M^2}{r^4}\Big(x^j\delta_{ik} + x^k\delta_{ij} - \frac{2x^i x^j x^k}{r^2}\Big),\nn\\
&&\stackrel{5~~~}{\Gamma_{j4}^i} = -\frac{3M}{2r}(\zeta_{i,j}-\zeta_{j,i})-\frac{M}{2r^3}(x^j\zeta_i+x^i\zeta_j). \nn
\end{eqnarray}

Following Schiff's approach \cite{Schiff1960a}, we can define the purely spatial vector $\bm{S}$ as
\begin{equation}\label{s-spin}
    \bm{S} = (S_1, S_2, S_3) \equiv (S^{23}, S^{31}, S^{12})~.
\end{equation}
Plugging Eq.~(\ref{Christoffel}) into Eq.~(\ref{sij}), and making use of Eq.~(\ref{s-spin}), we can obtain
\begin{eqnarray}
    \frac{d\bm{S}}{dt} &=& \frac{M}{r^3}\big[\left(\bm{x}\cdot\bm{S}\right)\bm{v} - 2\left(\bm{v}\cdot\bm{S}\right)\bm{x} + 3(\bm{v}\cdot\bm{x})\bm{S}\big] + \frac{1}{2}\bm{S}\times(\nabla\times\bm{\zeta})~.\label{spinEq}
\end{eqnarray}
When the gravitational source's rotation vanishes, Eq.~(\ref{spinEq}) reduces to the result of Schiff \cite{Schiff1960a}.

It is worth emphasizing that the spinning particle does not move along the metric geodesic, and Eq.~(\ref{spinEq}) is different from the spin evolution given by {\tt Eq.(9.6.5)} in the textbook \cite{Weinberg1972}, in which the spinning particle is treated as a point particle following the geodesic.

Following the same procedure given by Schiff, we can obtain the spin evolution measured in the rest frame of the particle as follows (see Appendix):
\begin{eqnarray}
\frac{d\bm{S}_{\text{rest}}}{dt}= \Omega\times\bm{S}_{\text{rest}}~,\label{precession_rest}
\end{eqnarray}
with
\begin{eqnarray}\label{Omega}
   \Omega = -\frac{1}{2}\nabla\times\bm{\zeta} -\frac{3M}{2r^3}(\bm{v}\times\bm{x})~.
\end{eqnarray}
The first term in the right hand side of Eq.~(\ref{Omega}) is just Lense-Thirring precession, and the second one is called as the geodetic precession or Thomas precession caused by gravitation~\cite{Weinberg1972}.

Although the formula for the precession frequency $\Omega$ is the same as {\tt Eq.(9.6.12)} in the textbook \cite{Weinberg1972}, both the equations of motion for the spinning particle and the evolution of the spin here differ from those in the textbook. These differences are due to the fact that we start with the Mathisson-Papapetrou equations, which are basing on the dipole approximation, to derive the equations of motion of the spinning particle and the spin evolution, while in the textbook the single-pole approximation is used to obtain the corresponding equations.

We can see from Eq.~(\ref{precession_rest}) that $\bm{S}_{\text{rest}}$ precesses at a rate $|\Omega|$ around the direction of $\Omega$. This implies that the magnitude of spin measured in the rest frame of the particle does not change with time in the post-Newtonian approximation, though the spinning particle does not move along the geodesic.

Now we turn to derive the equation of motion for the center-of-mass of the particle in Kerr spacetime. When taking into account of the effects of spin on the particle's motion, we only keep the coefficient of $S^{ij}$ on the right-hand side of Eq.~(\ref{motionEq:b}) to the next-to-leading terms. For the non-relativistic case in which the particle is assumed to be bounded by the gravitational source, Eq.~(\ref{motionEq:b}) can be approximated as
\begin{eqnarray}
    m \frac{Du^i}{Ds} &=& -\overset{2~~~~}{\Gamma_{j4,k}^4} u^k S^{ij} -\bigg(\overset{2~~~~}{\Gamma_{kl,j}^i} v^l + \overset{3~~~~}{\Gamma_{k4,j}^i}\bigg) u^4 S^{jk} + \bigg(\overset{3~~~}{\Gamma_{jk}^4}\overset{2~~~}{\Gamma_{44}^k} + \overset{2~~~}{\Gamma_{k4}^4}\overset{3~~~}{\Gamma_{j4}^k} - \overset{4~~~~}{\Gamma_{j4,k}^4}v^k \nn\\
    && - \overset{2~~~}{\Gamma_{j4}^4}\overset{2~~~}{\Gamma_{k4}^4}v^k + \overset{2~~~}{\Gamma_{k4}^4}\overset{2~~~}{\Gamma_{jl}^k}v^l - \overset{3~~~~}{\Gamma_{jl,k}^4}v^k v^l\bigg)u^4 S^{ij} - \bigg(\overset{5~~~~}{\Gamma_{k4,j}^i} + \overset{2~~~}{\Gamma_{jl}^i}\overset{3~~~}{\Gamma_{k4}^l} + \overset{3~~~}{\Gamma_{j4}^i}\overset{2~~~}{\Gamma_{k4}^4} \nn\\
    &&+ \overset{4~~~~}{\Gamma_{kl,j}^i} v^l + \overset{2~~~}{\Gamma_{jn}^i}\overset{2~~~}{\Gamma_{kl}^n} v^l + \overset{2~~~~}{\Gamma_{k4,l}^4}v^i u^j u^l\bigg)u^4 S^{jk}~,\label{eq:Du-i-low}\\
    m \frac{Du^4}{Ds} &=& -\overset{2~~~~}{\Gamma_{k4,j}^4}u^4 S^{jk} - \bigg(\overset{4~~~~}{\Gamma_{k4,j}^4} + \overset{3~~~~}{\Gamma_{kl,j}^4}v^l + \overset{2~~~~}{\Gamma_{k4,l}^4}u^j u^l\bigg)u^4 S^{jk}~.\label{eq:Du-0-low}
\end{eqnarray}
Substituting Eqs.~(\ref{eq:Du-i-low}) and (\ref{eq:Du-0-low}) into Eq.~(\ref{eq:acceleration}) and making the approximation $1/u^4\approx1-M/r-v^2/2$, we can obtain the acceleration of the non-relativistic particle as follow
\begin{eqnarray}
\frac{d\bm{v}}{d t} &=& \bm{a}_\mathrm{low} + \frac{3M}{m r^5}\Big(1 -\frac{2M}{r} - \frac{1}{2}v^2\Big) \big\{[\bm{x}\cdot(\bm{v}\times\bm{S})]\bm{x} -r^2(\bm{v} \times \bm{S}) + 2(\bm{v}\cdot\bm{x}) (\bm{x}\times\bm{S})\big\} \nn\\
&& + \frac{1}{2 m}\Big(1 -\frac{4M}{r} - \frac{1}{2}v^2\Big)\nabla [\bm{S}\cdot(\nabla\times\bm{\zeta})] + \frac{3M}{2m r^5}[\bm{x}\cdot(\bm{S}\times\bm{\zeta})]\bm{x} \nn\\
&& -\frac{1}{2m}\Big\{(\bm{v}\cdot\nabla)^2(\bm{S}\times\bm{\zeta}) + (\bm{v}\cdot\nabla)(\bm{S}\times\nabla)(\bm{v}\cdot\bm{\zeta}) + [\bm{v}\cdot\nabla(\bm{S}\cdot(\nabla\times\bm{\zeta}))]\bm{v}\Big\} \nn\\
&& + \frac{M}{2mr^3}\Big\{\bm{S}\times\bm{\zeta} -\bm{S}\!\times\!\nabla(\bm{x}\cdot\bm{\zeta}) - \nabla[\bm{x}\cdot(\bm{S}\times\bm{\zeta})] + 2[\bm{x}\cdot(\bm{S}\times\nabla)]\bm{\zeta} \nn\\
&&  + 3(\bm{x}\cdot\nabla)(\bm{S}\times\bm{\zeta}) + 3[\bm{S}\cdot(\nabla\times\bm{\zeta})]\bm{x}\Big\},\label{lowaccer}
\end{eqnarray}
where $\bm{a}_\mathrm{low}$ represents the acceleration of the non-relativistic particle along the geodesic of the Kerr spacetime, which can be found in our previous work~\cite{JiangLin2014}.
We have also verified that when the reference worldline of the particle is shifted to $\tilde{x}^i = x^i -S^{ij}u_j/m$, this equation confirms the relative one-body equation of motion in a two-body system~\cite{Kidder1995} in the extreme-mass-ratio limit (see {\tt Eqs.(2.1) and (2.2)} with $\eta \rightarrow 0$ therein), but our result includes higher-order effects of the spin.

For the spinning particle with a relativistic velocity, we have $O(v^i)\sim 1$. The evolution equations of spatial components of $S^{\alpha\beta}$ can be derived from Eq.~(\ref{sij}), which reads
\begin{eqnarray}
  \frac{d\bm{S}}{dt} &=& \frac{M}{r^3}\big[\left(\bm{x}\cdot\bm{S}\right)\bm{v} - 2\left(\bm{v}\cdot\bm{S}\right)\bm{x} + 3(\bm{v}\cdot\bm{x})\bm{S}\big] + \frac12 \bm{S}\times(\nabla\times\bm{\zeta}) \nn\\
  &&- [\bm{v}\cdot\nabla(\bm{v}\cdot\bm{\zeta})]\bm{S} + \frac12 (\bm{v}\cdot\bm{S})\big[(\bm{v}\cdot\nabla)\bm{\zeta} + \nabla(\bm{v}\cdot\bm{\zeta})\big]~.\label{eq:spin-high}
\end{eqnarray}
The equation of motion of the relativistic particle can be derived from Eq.~(\ref{motionEq:b}), which can be approximated as
\begin{eqnarray}
    m \frac{Du^i}{Ds} \!&=&\! -\bigg(\stackrel{2~~~~}{\Gamma_{j4,k}^4} +\stackrel{3~~~~}{\Gamma_{jl,k}^4}v^l\bigg) u^k S^{ij} \nn\\
    &&-\bigg(\stackrel{2~~~~}{\Gamma_{kl,j}^i} v^l +\stackrel{2~~~~}{\Gamma_{k4,l}^4} v^i u^j u^l +\stackrel{3~~~~}{\Gamma_{k4,j}^i} +\stackrel{3~~~~}{\Gamma_{kn,l}^4}v^i v^j u^l u^n\bigg)u^4 S^{jk}, \label{eq:Du-i-high}\\
    m \frac{Du^4}{Ds} \!&=&\! - \bigg(\!\stackrel{2~~~~}{\Gamma_{k4,j}^4} + \stackrel{3~~~~}{\Gamma_{kl,j}^4}v^l + \stackrel{2~~~~}{\Gamma_{k4,l}^4}u^j u^l +\stackrel{3~~~~}{\Gamma_{kn,l}^4}u^j u^n u^l\bigg)u^4 S^{jk}. \label{eq:Du-0-high}
\end{eqnarray}
Substituting Eqs.~(\ref{eq:Du-i-high}) and (\ref{eq:Du-0-high}) into Eq.~(\ref{eq:acceleration}), we can obtain the dynamics for the relativistic particle with spin as follow
\begin{eqnarray}
&& \hskip -1cm \frac{d\bm{v}}{dt} = \bm{a}_\mathrm{high} + \frac{3M}{\gamma m  r^5}\{[\bm{x}\cdot(\bm{v}\times\bm{S})]\bm{x} -r^2(\bm{v}\times\bm{S})+2(\bm{v}\cdot\bm{x})(\bm{x}\times\bm{S})\} \nn \\
&& -\frac{1}{2\gamma m}\Big\{(\bm{v}\cdot\nabla)^2(\bm{S}\times\bm{\zeta}) + (\bm{v}\cdot\nabla)(\bm{S}\times\nabla)(\bm{v}\cdot\bm{\zeta}) + [\bm{v}\cdot\nabla(\bm{S}\cdot(\nabla\times\bm{\zeta}))]\bm{v} \nn\\
&& - \nabla [\bm{S}\cdot(\nabla\times\bm{\zeta})]\Big\},\label{highaccer}
\end{eqnarray}
where $\bm{a}_\mathrm{high}$ is the acceleration of the relativistic particle along the geodesic of the Kerr spacetime and has been derived in Ref.~\cite{JiangLin2014}. $\gamma = (1-v^2)^{-\frac{1}{2}}$ is Lorentz factor. It is observed that the effects of spin-orbit and spin-spin couplings are suppressed by a factor of $\gamma$ in the relativistic case.

\section{Summary}\label{conclusion}

The motion of the spinning particle under the CP condition in the Kerr spacetime has been investigated. We derive the post-Newtonian equation for the evolution of spin, and the acceleration of the spinning particle due to the coupling of spin-curvature to the next-to-leading terms. We also calculate the precession of spin measured in the rest frame of the particle when the particle is non-relativistic, and find that the measured spin's magnitude does not change with time though the spinning particle does not move along the metric geodesic.
Moreover, it is found that the effects of the spin-orbit and spin-spin couplings in the equations of motion are suppressed by the Lorentz factor when the spinning particle has a relativistic velocity.

\appendix
\section{The evolution of spin in the rest frame of the particle.}\label{app}

Let $\bm{S}_{\text{rest}}$ denote the spin with respective to the rest frame of the particle. When the particle's velocity $\bm{v}$ is non-relativistic, we have \cite{Schiff1960a}
\begin{eqnarray}\label{transformation}
    \bm{S}_{\text{rest}} = \Big(1+\frac{2M}{r}\Big)\bm{S} + \frac{1}{2}\left[v^2\bm{S}-\left(\bm{v}\cdot\bm{S}\right)\bm{v}\right]~,
\end{eqnarray}
where $v \equiv |\bm{v}|$.

Differentiating this equation with the time $t$, we can get
\begin{eqnarray}
\frac{d\bm{S}_{\text{rest}}}{dt} &=& \frac{1}{2}\bigg[2v\frac{dv}{dt}\bm{S}+ v^2\frac{d\bm{S}}{dt} - \frac{d\bm{v}}{dt}(\bm{v} \cdot \bm{S}) - \bm{v}\left(\frac{d\bm{v}}{dt}\cdot\bm{S}\right) - \bm{v}\left(\bm{v}\cdot\frac{d\bm{S}}{dt}\right)\bigg] \nn\\
&&-\frac{2M}{r^2}\frac{dr}{dt}\bm{S} + \left(1+\frac{2M}{r}\right) \frac{d\bm{S}}{dt}~.\label{dSrestdt}
\end{eqnarray}
From Eq.~(\ref{spinEq}) we have
\begin{eqnarray}\label{dSdtO}
\frac{d\bm{S}}{dt} = O\left(\frac{v^3}{r}\right)\bm{S} = O\left(\frac{v^3}{r}\right)\bm{S}_{\text{rest}}~,
\end{eqnarray}
Making use of Eq.~(\ref{dSdtO}) and the following relations
\begin{eqnarray*}
&&dr/dt=(\bm{x}\cdot\bm{v})/r, \quad d\bm{v}/dt \approx-M\bm{x}/r^3, \\
&&dv/dt=(\bm{v}\cdot d\bm{v}/dt)/v~,
\end{eqnarray*}
we can obtain the rate of change of $\bm{S}_{\text{rest}}$ to order $(v^3/r)\bm{S}_{\text{rest}}$ from Eq.~(\ref{dSrestdt}) as follows
\begin{eqnarray*}
\frac{d\bm{S}_{\text{rest}}}{dt} &=& \frac{d\bm{S}}{dt} + \frac{M}{2r^3}\left[(\bm{v} \cdot \bm{S})\bm{x} + (\bm{x} \cdot \bm{S})\bm{v} -6(\bm{v}\cdot\bm{x})\bm{S}\right]~.
\end{eqnarray*}
Plugging Eq.~(\ref{spinEq}) into the above equation, we have
\begin{eqnarray}
\frac{d\bm{S}_{\text{rest}}}{dt}
&\approx& \frac{1}{2}\bm{S}\times(\nabla\times\bm{\zeta}) +\frac{3M}{2r^3}[(\bm{x} \cdot \bm{S})\bm{v} -(\bm{v} \cdot \bm{S})\bm{x}] \nn\\
&=&\bm{S}\times\left[\frac{1}{2}\nabla\times\bm{\zeta} +\frac{3M}{2r^3}(\bm{v}\times\bm{x})\right] \nn\\
&\approx& \underbrace{-\left[\frac{1}{2}\nabla\times\bm{\zeta} +\frac{3M}{2r^3}(\bm{v}\times\bm{x})\right]}_{\Omega}\times\bm{S}_{\text{rest}}~,
\end{eqnarray}
where $\Omega$ represents the precession frequency.

\acknowledgments

This work was supported in part by the Ph.D. Program Foundation of Ministry of Education of China (No. 20110184110016) and the National Basic Research Program of China (973 Program, No. 2013CB328904), as well as the Fundamental Research Funds for the Central Universities (No. 2682014ZT32).

\end{document}